# Trapping and localization of particles by a potential well deepening with time

Azad Ch. Izmailov

Institute of Physics, Azerbaijan National Academy of Sciences, Javid av. 33, Baku, Az-1143, AZERBAIJAN

*e-mail*: azizm57@rambler.ru

## *Abstract*

The universal mechanism of trapping and localization of sufficiently slow-speed particles by a potential well deepening with time is established on the basis of fundamental relations of classical mechanics. Such wells may be created for a wide range both charged and neutral particles (with electric or magnetic moments) by means of a controllable electromagnetic field with growing strength (up to a certain moment) but with a fixed spatial distribution. Detailed analysis of features of corresponding electromagnetic traps of particles is carried out on the visual example of the one-dimensional rectangular potential well deepening with time. Obtained results may be used in ultrahigh resolution spectroscopy of various microparticles (including atoms, molecules and ions).

## 1. Introduction

Electromagnetic traps for free charged and neutral particles without material walls allow to localize and observe these particles during a comparatively long period of time thereby creating conditions for detailed research of their properties [1]. In particular, such traps of microparticles in the high vacuum open new possibilities for contactless measurements of forces acting on given particles with extremely high accuracy and allow micromanipulations of these particles [2]. Even more important is the development of effective methods of trapping and localization of sufficiently slow-speed atoms, molecules and ions for ultrahigh resolution spectroscopy [3] and for creation of more precise standards of time and frequency [4].

In the short author's preprint [5], the idea was stated on possible trapping and localization of particles in a potential well induced by an electromagnetic field increasing with time (up to a certain moment). Corresponding calculations in this work [5] were carried out on example of the one-dimensional classical oscillator. Depending on whether particles have electric (or magnetic) moment, it is possible to use the controllable electric (or magnetic) field for such their trapping.

In the present work we establish the universality of this mechanism of trapping and localization of sufficiently slow-speed particles on the basis of fundamental relations of classical mechanics (section 2). Detailed analysis of features of corresponding electromagnetic traps of particles is carried out on the simple example of the one-dimensional rectangular well deepening with time (section 3).

## 2. Basic relationships

Let us consider a point particle with the mass *m* freely moving in a three-dimensional space before its entering to the region *V* of the potential well $U(r,t)$, which explicitly depends not only on the coordinate $r$ but also on time $t$. The total energy of such a particle with the non-relativistic velocity $v$ is described by the known formula [6]:

$$E(r,v,t) = 0.5mv^2 + U(r,t). \qquad (1)$$

Further we will consider the potential energy $U(r,t)$ of the following form:

$$U(r,t) = \sigma(r) * \varphi(t), \qquad (2)$$

where the coordinate function $\sigma(r) \leq 0$ in the region V, and $\varphi(t) \geq 0$ is the nondecreasing function of time $t$. Such a potential well (2) for particles may be created by a controllable electromagnetic field with the growing strength (up to a certain time moment) but with a fixed spatial distribution [7]. In this case the motion equation [6] for the particle has the form:

$$m \frac{d^2 r}{dt^2} = -\varphi(t) \frac{d\sigma(r)}{dr}. \qquad (3)$$

On the basis of relationships (1)-(3) we directly receive the following formula for the time derivative of the total energy $E(r,v,t)$:

$$\frac{dE}{dt} = \sigma(r) \frac{d\varphi(t)}{dt} \leq 0. \qquad (4)$$

Thus, according to (4), an increase of the function $\varphi(t)$ with time $t$ leads to the decrease of the total energy $E(r,v,t)$ (1) of the particle in the region V of the potential well, where $\sigma(r) < 0$. We see also from the formula (1) that the particle can not go beyond the potential well (where $U(r,t) = 0$), when its energy $E$ will be negative. It is important to note that such a classical particle will be localized in the region V of the potential well even after output of the nondecreasing function $\varphi(t)$ on a constant maximum value, when a negative total energy $E < 0$ of this particle will be conserved. It is obvious, that sufficiently fast particles will not be captured in the considered trap. Detailed analysis of dynamics of particles may be carried out for electromagnetic traps with definite spatial configurations. In the next section we will establish a number of important features of trapping and localization of sufficiently slow-speed non-relativistic particles on the visual example of the one-dimensional rectangular potential well.

### 3. One-dimensional rectangular potential well

Let us consider a point particle with the mass $m$ freely moving with the velocity $v_0 > 0$ along the axis $x$ (Fig.1) from the region $x < -L$ and in a certain moment $t$ reaching the boundary $x = -L$ of the following potential well:

$$U(x,t) = -J_0 * \varphi(t) * \eta(L^2 - x^2), \qquad (5)$$

where $J_0 > 0$ is the constant value with the energy dimension, $1 \geq \varphi(t) \geq 0$ is the nondecreasing function of time $t$, $\eta(y)$ is the step function ($\eta(y) = 1$ for $y \geq 0$ and $\eta(y) = 0$ if $y < 0$). From (3) we receive the motion equation of the particle for the potential well (5):

$$m \frac{d^2 x}{dt^2} = J_0 * \varphi(t) * [\delta(x+L) - \delta(x-L)], \qquad (6)$$

where $\delta(y)$ is the Dirac delta-function. According to Eq.(6), an abrupt increase of the particle velocity occurs from the initial value $v_0$ to $v \geq v_0$, when this particle falls into the well (5) in the moment $t$. Connection between given values $v_0$ and $v$ is determined from the formula (1) for the total energy of the classical particle in this moment $t$:

$$E(-L, t) = 0.5mv_0^2 = 0.5mv^2 - J_0 * \varphi(t). \tag{7}$$

According to Eq.(6), the considered particle further will move inside the rectangular well (5) with the constant velocity $v$ and will reach the opposite well boundary (with the coordinate $x = L$) in the moment $(t + 2L/v)$. However this particle will not be able to overcome the potential well if its total energy $E$ (1) will become negative because of an increase of the function $\varphi(t)$ (7) with time $t$, that is at the following condition:

$$0.5mv^2 - J_0 * \varphi(t + 2L/v) < 0. \tag{8}$$

Under condition (8), the particle is reflected from the well boundary with the coordinate $x = L$ and will move in the reverse direction with the constant velocity $(-v)$ up to arrival to the opposite well boundary (with the coordinate $x = -L$) in the moment $(t + 4L/v)$. Because of the relationship $\varphi(t + 4L/v) \geq \varphi(t + 2L/v)$, the similar motion of the particle with the velocity $v$ from the boundary $x = -L$ of the potential well up to $x = L$ will be repeated and so on. Thus, according to condition (8), the maximum possible speed $v_{max}(t)$ of particles, captured in the considered trap (5) in the moment $t$, is determined by the equation:

$$0.5mv_{max}^2 = J_0 * \varphi(t + 2L/v_{max}). \tag{9}$$

After finding of the value $v_{max}(t)$ from Eq.(9), we obtain from formula (7) the maximum speed $\tilde{v}_0(t)$ of a free particle, which, after falling into the potential well (5) in the moment $t$, will be localized in this well:

$$\tilde{v}_0(t) = \sqrt{v_{max}^2(t) - \frac{2J_0 \, \varphi(t)}{m}}. \tag{10}$$

The minimum speed $v_{min}(t)$ of particles, captured in the considered trap in the moment $t$, is determined from formula (7) at the value $v_0 = 0$:

$$v_{min}(t) = \sqrt{\frac{2J_0 \, \varphi(t)}{m}}. \tag{11}$$

Let us assume, that the function $\varphi(t)$ in (5) increases up to the maximum value 1 during the period from 0 to $T$ and then will be constant, that is $\varphi(t) = 1$ if $t \geq T$. In this connection we introduce following characteristic values for the potential well (5) with dimensions of speed and energy:

$$w = 2L/T, \quad K = 0.5mw^2. \tag{12}$$

In case of the comparatively shallow well (5), when $J_0 \leq K$, we receive from Eq.(9) the maximum speed $v_{max}(t)$ of particles captured in the trap, which is constant during the period $0 \leq t \leq T$ of growth of the function $\varphi(t)$:

$$v_{max}(t) = v^* = \sqrt{\frac{2J_0}{m}} = w * \sqrt{\frac{J_0}{K}}. \tag{13}$$

Then we obtain the corresponding allowed initial speed of a free particle from (10):

$$\tilde{v}_0(t) = \sqrt{\frac{2J_0}{m}[1 - \varphi(t)]}, \quad (0 \leq t \leq T). \tag{14}$$

In case of a sufficiently large depth of the potential well (5), when $J_0 > K$, according to Eq.(9), the speed $v_{max}(t)$ of particles, captured in the trap, reaches the constant maximum value $v^*$ (13) in the time period $t^* \leq t \leq T$, where

$$t^* = T - 2L * \sqrt{\frac{m}{2J_0}} = \left(1 - \sqrt{\frac{K}{J_0}}\right) T. \qquad (15)$$

Further, for definiteness, we consider the following time dependence $\varphi(t)$ of the potential (5):

$$\varphi(t) = \left(\frac{t}{T}\right)^n \eta(T - t), \qquad (n > 0, \ t \geq 0). \qquad (16)$$

According to formula (16), the depth of the potential well (5) increases from 0 to $J_0$ during the period $0 \leq t \leq T$, and will have the maximum constant value $J_0$ when $t > T$.

Fig.2 presents dependences of speeds $\tilde{v}_0(t)$, $v_{max}(t)$ and $v_{min}(t)$ (9)-(11) of trapped particles on time $t$ for 2 parameters $n=0.5$ and 2 of the given function $\varphi(t)$ (16). We see that the considered potential well (5) carries out trapping and localization of free particles, whose speeds $|v_0|$ in a moment $t$ of its falling in the well is restricted by the value $\tilde{v}_0(t) > |v_0|$ from relationships (9), (10). Then the speed $|v|$ of given captured particles inside the well is between values $v_{max}(t)$ and $v_{min}(t)$ (Fig.2) determined by (9), (11). Trapping of new particles in the potential well (5), (16) stops after the moment $T$, when this well will be stationary (Fig.2). However classical particles, captured in the given electromagnetic trap before the moment $T$, will remain there also in following time $t > T$. Possible speeds $|v|$ of these particles, remained inside the potential well, will be between 0 and $\sqrt{2J_0/m}$. According to Fig.2, speed intervals of particles, captured in such an electromagnetic trap in fixed moments $t < T$, essentially depend on the parameter $n$ of the function $\varphi(t)$ (16), that is on increase rate of the well depth with time $t$. At the sufficiently shallow potential well (5), when $J_0 \leq K$, corresponding speeds of particles in Fig.2a,b are described by simple formulas $v_{min}(t)$ (11), $v_{max}(t) = v^*$ (13) and $\tilde{v}_0(t)$ (14). In case $J_0 > K$, increase of the speed $v_{max}(t)$ occurs in the time period $0 \leq t \leq t^*$ (15) and the constant value $v_{max}(t) = v^*$ (13) establishes after the moment $t^*$ (Fig.2c,d). In particular, for the given function $\varphi(t)$ (16), speeds $\tilde{v}_0(t)$ and $v_{max}(t)$ coincide in the initial moment $t = 0$ and, according to (9), (10), have the following form when $J_0 \geq K$:

$$v_{max}(t=0) = \tilde{v}_0(t=0) = \left(\frac{2J_0}{m}\right)^{\frac{1}{(n+2)}} \left(\frac{2L}{T}\right)^{\frac{n}{(n+2)}} = \left(\frac{J_0}{K}\right)^{\frac{1}{(n+2)}} w. \qquad (17)$$

According to (17), available speeds of trapped particles increase with growth of the length $2L$ and the depth $J_0$ of the potential well (5), (16) and also with decrease of the characteristic time $T$ of its deepening.

The considered one-dimensional rectangular potential well (5) may be created, for example, by the controllable local homogeneous electric (or magnetic) field on the propagation path of a collimated beam of classical particles having electric (or magnetic) moment. During the definite time $T$ of growth of the given field strength (up to a certain maximum value), such an electromagnetic well will capture sufficiently slow-speed particles from the beam. Characteristic speeds of these trapped particles may be estimated on the basis of relationships obtained in the present work.

## 4. Conclusions

From fundamental relations of classical mechanics, we have shown possibility of trapping and localization of sufficiently slow-speed particles by a potential well deepening with time, which is described by the general formula (2). Such universal traps may be created in practice for various charged and neutral classical particles (having electric or magnetic moments) by means of the controllable electric or magnetic field with strength increasing during a certain period but at a fixed spatial field distribution. Similar potential wells may be induced also by amplifying laser beams of definite spatial configurations.

In section 3 we have established a number of interesting features of such electromagnetic traps on the visual example of the one-dimensional rectangular potential well, which confirm basic conclusions of the previous section 2. Thus, even a highly shallow but increasing with time potential wells will continuously capture sufficiently slow-speed particles. Such trapped classical particles will remain in the potential well even after going out of the corresponding nondecreasing electromagnetic field on a stationary value.

Of course, results obtained in the present work are valid only in the absence of an interaction between particles travelling through a potential well. However such an interaction may be essential at a sufficiently high concentration of captured particles in a comparatively small volume of an electromagnetic trap.

It should be noted that dynamics of classical and quantum systems in time-dependent potentials was investigated earlier in a series of important publications, including books [8,9]. However these earlier papers did not contain analysis of the universal mechanism of trapping and localization of classical particles considered in the present work. At the same time, basic mathematical relationships obtained in given previous articles may be used in the future for detailed research of proposed nonstationary electromagnetic traps with various spatial configurations, taking into account their concrete important applications.

**References**

[1]  W. Paul, *Electromagnetic traps for charged and neutral particles* (Nobel Lecture), Angewandte Chemie, International Edition in English, 1990, V.29, issue 7, pages 739-748.

[2]  A. Ashkin, *Optical Trapping and Manipulation of Neutral Particles Using Lasers*. (World Scientific

Publishing, 2006).

[3]  W. Demtroder, *Laser Spectroscopy: Basic Concepts and Instrumentation*. (Springer, Berlin, 2003).

[4]  F. Riehle, *Frequency Standards-Basics and Applications*. (Berlin: Wiley-VCH, 2004).

[5]  A.Ch. Izmailov, *Trapping of slow-speed particles in a gas cell by the nonhomogeneous electromagnetic field intensifying with time*, arXiv: 1401.2718, January 2014.

[6]  L.D. Landau and E.M. Lifshitz, *Mechanics. Course of Theoretical Physics. Volume 1*. (Elsevier Ltd, 1976).

[7]  M. Mansuripur, *Field, Force, Energy and Momentum in Classical Electrodynamics* (Bentham e Books, 2011).

[8]  P.G.L. Leach, S.E. Bouquet, J.-L. Rouet, and E. Fijalkow, *Dynamical Systems, Plasmas and Gravitation.*, Selected papers from the Conference held in Orleans la Source, France, 22-24 June 1997 (Springer, 1999).

[9]  A. Mostafazadeh, *Dynamical invariants, adiabatic approximation and the geometric phase* (Nova Science Publishers, 2001).

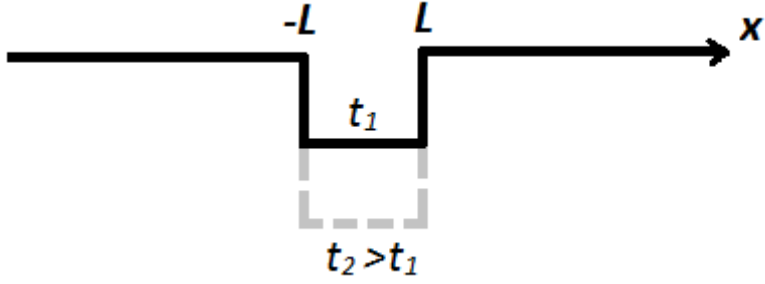

**Fig.1.** Scheme of the one-dimensional rectangular potential well deepening with time *t*.

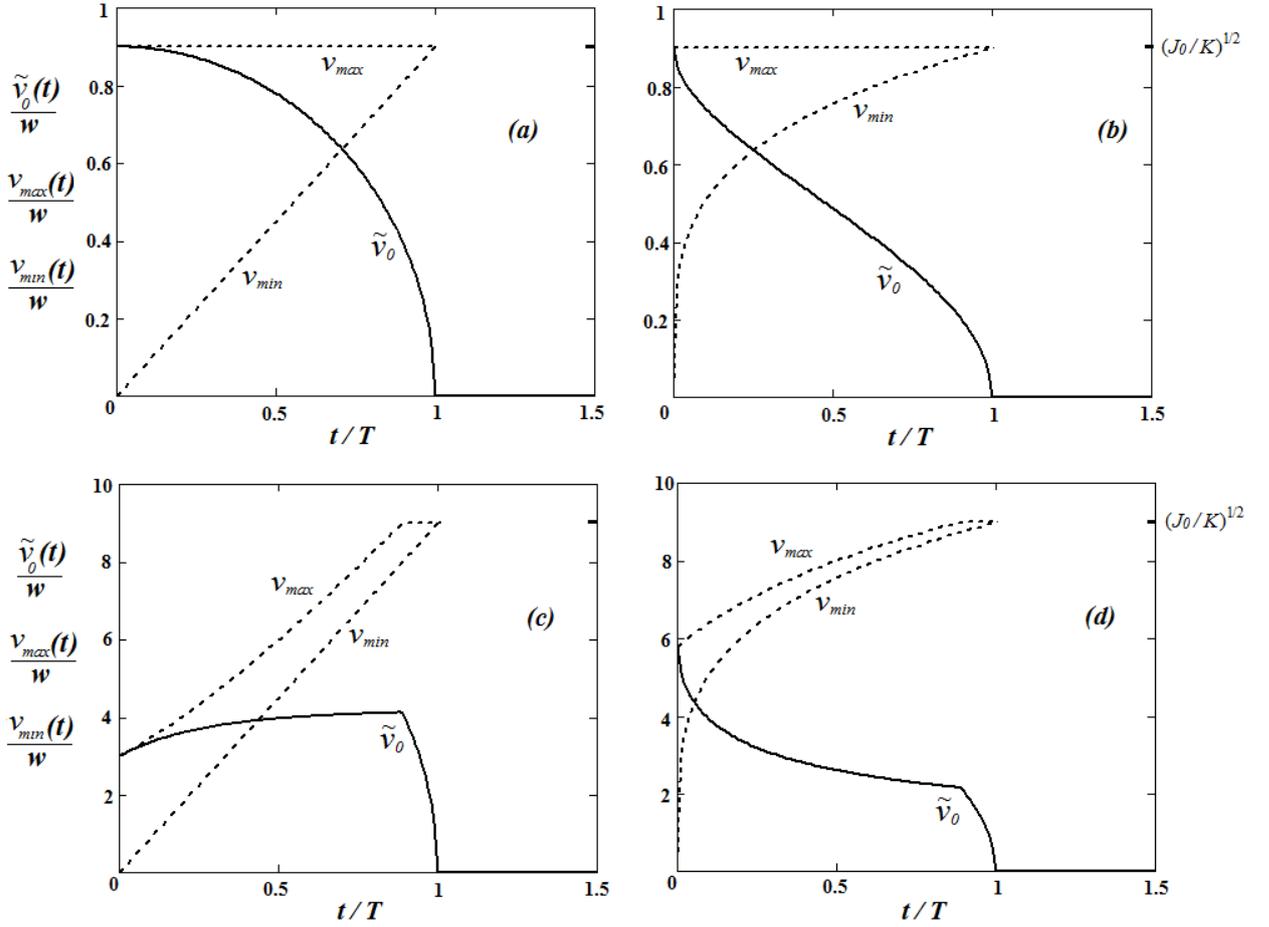

**Fig.2.** Speeds $\tilde{v}_0(t)$, $v_{max}(t)$ and $v_{min}(t)$ of trapped particles versus time $t$ of their getting into the one-dimensional rectangular potential well (5) with the value $J_0 = 0.81K$ (a, b) and $81K$ (c,d) for the parameter $n = 2$ (a, c) and 0.5 (b, d) of the function $\varphi(t)$ (16).

6